\documentclass[proceedings]{ccn}  
\ccndoi{10.32470/04ojkmgu}  
\title{Topographic constraints shape brain-like component structure in auditory models}

\author{%
  Haider Al-Tahan\affmark{1,2} \And
  Mayukh Deb\affmark{1,2} \And
  Jenelle Feather\affmark{3,4,5} \And
  N. Apurva Ratan Murty\affmark{1,2} 
}
\affiliation{1}{School of Psychological and Brain Sciences, Georgia Tech}
\affiliation{2}{Center for Computational Cognition, School of Psychological and Brain Sciences, Georgia Tech}
\affiliation{3}{Center for Computational Neuroscience, Flatiron Institute, Simons Foundation}
\affiliation{4}{Psychology Department, Carnegie Mellon University}
\affiliation{5}{Neuroscience Institute, Carnegie Mellon University}
\emails{{haideraltahan, mayukh, ratan}@gatech.edu; jfeather@cmu.edu}
\newcommand{\toponame}{TopoAudio}
\newcommand{\toponames}{TopoAudio }

\usepackage{caption}
\usepackage{booktabs}
\usepackage{placeins} 

\usepackage{booktabs}
\usepackage{subcaption} 
\usepackage{colortbl}
\usepackage{xcolor}
\usepackage{cleveref}
\usepackage{adjustbox}

\begin{document}

\twocolumn[{%
\renewcommand\twocolumn[1][]{#1}%
\maketitle

\begin{center}
  \centering
  \captionsetup{type=figure}
  \includegraphics[width=0.8\linewidth]{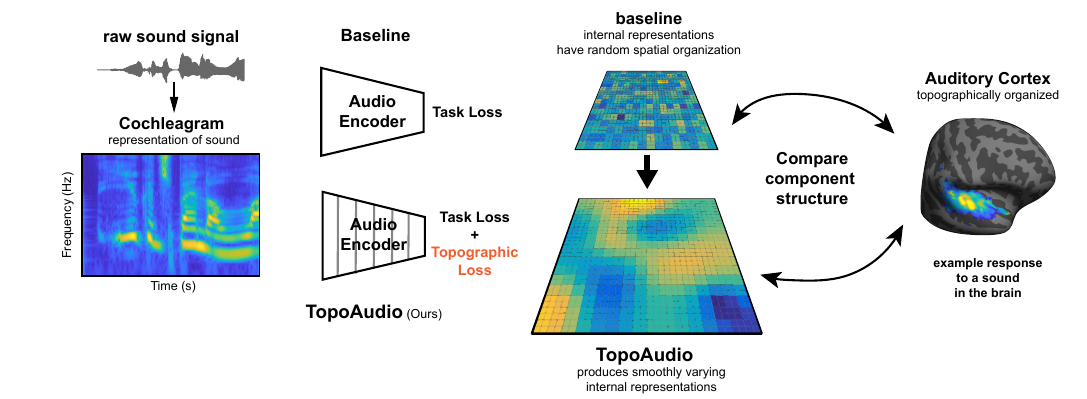}
  \captionof{figure}{\textbf{\toponame: Topographic auditory models.} Raw sound waveforms are transformed into cochleagrams and passed through the neural network backbones (middle). The baseline model (middle, top) is trained using only a task loss. In contrast, TopoAudio is trained with an additional spatial smoothness constraint via TopoLoss. In this work we compare the component structure of the internal representations of baseline and topographic audio models to those inferred from the human auditory cortex (right)}
  \label{fig:summary}
\end{center}
}]

\begin{abstract}
  If topography is a fundamental feature of the brain, it should influence both how neurons are arranged in space (i.e. explain brain maps) and how information is structured within the neural population. The human auditory cortex provides a strong, but previously underused test for the latter idea. Neural responses measured with both fMRI and ECoG can be decomposed into interpretable components corresponding to sound categories such as speech, music, and song, offering a view of how sound information is partitioned in the brain. Here we ask whether introducing topographic constraints into the training of audio neural network models shapes their internal representations to better match the component structure observed in the brain. To address this question, we introduce a new class of topographic auditory models, TopoAudio, which incorporate wiring-length constraints and encourage nearby units on a two-dimensional cortical sheet to develop similar response tuning. Despite these additional constraints we find that TopoAudio achieves comparable performance on standard speech and environmental sound classification tasks to standard non-topographic models and matches them in predicting human fMRI responses. Crucially however, topographic models develop more compact internal representations, and their inferred components align more closely with those derived from human ECoG recordings. These results provide initial evidence that topography offers a general mechanism for producing biologically aligned internal representations in artificial neural networks. More broadly, component-level alignment provides a complementary way for testing whether topography reshapes population responses in models to better match the representational structure observed in neural recordings. Our project page: \href{https://topoaudio.github.io/}{topoaudio.github.io}\footnote{This work was supported in part by the NIH Pathway to Independence Award (R00EY032603), NSF Nexus (Allocation number: SOC250049), a startup grant from Georgia Tech to N.A.R.M., and an NSERC Postgraduate Scholarship – Doctoral. We thank everyone at Murty Lab for their feedback and support.} 
  
\end{abstract}

\section{Introduction}

Neural responses in the brain are topographically organized in many parts of cortex. Why such a topographic organization emerges in brains and how topography shapes the underlying neural code remains an open question. A leading hypothesis is that topography imposes a `wiring-length minimization' constraint on neurons which explains why neurons with similar response properties cluster together \citep{KAAS1997107, chklovskii2004maps, jacobs1992computational}. But if topography is such a core constraint for the brain it should influence not just the spatial organization of neurons (producing orderly brain maps etc.) but also explain how information is partitioned and represented \textit{within} the neural population. While spatial aspects of the neural organization have received a lot of attention, far less is known about whether (and if at all) topography shapes the structure of neural representations. 

Artificial neural networks (ANNs) offer a practical way to test some of these ideas. Multiple approaches have now been developed to induce topographic organization within commonplace AI architectures \citep{blauch2022connectivity, lee2020topographic, margalit_unifying_2024, qian2024local, deb2025toponets}. In vision, topography has been shown to induce spatial maps (pinwheels and high-level category-selective patches) and modestly improve predictivity on neural data \citep{lee2020topographic, margalit_unifying_2024, blauch2022connectivity, qian2024local, deb2025toponets}. Similar strategies have also been applied to induce topography into language models to study the spatial clustering of units within LLMs \citep{deb2025toponets, rathi2024topolm}.  But prior work has largely evaluated these models in terms of the emergence of spatially smooth maps or improvements in neural predictivity leaving open a deeper question: does topography fundamentally alter how information is internally represented within models?

As such, the auditory system offers a strong, but previously unexplored, test case to evaluate the effect of topography within ANNs. Prior work has shown that neural responses in the auditory cortex (as measured with fMRI and human ECoG) can be decomposed into a small set of core components (or dimensions) that capture how sound information is internally represented \citep{norman2015distinct}. Further, many of the component responses are directly interpretable and remarkably specific for complex acoustic features like speech and music \citep{norman2015distinct}, and even song \citep{norman2022neural}.  

In this work we ask whether imposing topography into audio-ANNs changes model representations such that their derived component structure \textit{better} matches the component structure observed in the human auditory cortex. We already know that introducing topography lowers the intrinsic dimensionality of model representations \citep{margalit_unifying_2024, deb2025toponets, qian2024local}. But whether the inferred dimensions (henceforth components) A) change because of induced topography or B) more closely align with the components inferred from the brain, has been hard to evaluate in vision and language models. This is largely because a comprehensive inventory of neural components has been lacking in vision and language.

To this end, we trained a suite of auditory ANN models (\toponame) on the same set of audio tasks, both with and without topographic constraints, enabling a controlled comparison of how topography changes internal representations. We first show that topographic models remain strong candidate models of auditory processing. They achieve high performance on standard audio classification benchmarks and predict human auditory fMRI responses as well as the non-topographic baseline model. We then turn to the central question of the paper: does imposing topography reshape the internal representation of \toponames in a brain-like way? 

We find that imposing topography changes internal representation of models by reducing the number of representational components and yielding a more compact organization. Critically, the components recovered from topographic models show stronger alignment with the component structure derived from human auditory cortex using ECoG, indicating that topography reshapes internal representations toward a more brain-like organization.

\section{Methods}
\label{sec:methods}

\subsection{Model training}
\subsubsection{Spatial Loss.}
To investigate how topographic constraints shape auditory representations, we adapted the TopoLoss framework \citep{deb2025toponets} to the auditory domain. As in prior studies on topography, we define a 2D "cortical sheet" from convolutional layers in the auditory model on which to enforce topography. Each convolutional kernel in the model is mapped onto this sheet. For a convolutional layer with $c_{\text{input}}$ input channels and $c_{\text{output}}$ output channels, and a kernel size of $k \times k$, the weight tensor $W \in \mathbb{R}^{c_{\text{output}} \times c_{\text{input}} \times k \times k}$ is reshaped into a cortical representation $C \in \mathbb{R}^{h \times w \times d}$, where $h \times w = c_{\text{output}}$, and $d = c_{\text{input}} \cdot k \cdot k$. 

To encourage smoothness in the cortical sheet $C^{h \times w \times d}$, we apply a blurring operation that removes high-frequency variations. We compute a blurred version $C'$ of the cortical sheet using a downsampling factor $\phi_h = \phi_w = 3$ followed by upsampling:

\begin{equation}
\text{Blur}(X, \phi_h, \phi_w) = f_{\text{up}}\left(f_{\text{down}}\left(X, \frac{h}{\phi_h}, \frac{w}{\phi_w} \right), h, w\right)
\end{equation}

The \textit{TopoLoss} is then defined as the negative mean cosine similarity between the original and blurred cortical maps:

\begin{equation}
\mathcal{L}_{\text{topo}} = -\frac{1}{N} \sum_{i=1}^N \frac{C_i \cdot C'_i}{\|C_i\| \|C'_i\|}
\end{equation}

This encourages neurons with similar functions to be spatially clustered, enhancing topographic organization. Finally, we integrate the \textit{TopoLoss} with the primary task loss $\mathcal{L}_{\text{training}}$ as:

\begin{equation}
\mathcal{L}_{\text{total}} = \mathcal{L}_{\text{training}} + \tau \cdot \mathcal{L}_{\text{topo}}
\end{equation}

where $\tau$ is a scaling coefficient controlling the influence of topographic regularization. Higher values of $\tau$ encourage stronger topographic organization.

\subsubsection{\toponames architecture and training.}
Our topographic models were derived from the popular auditory neural network CochResNet50. We selected this architecture because it was the most accurate model of human auditory cortex responses \citep{tuckute_many_2023}. CochResNet50 adapts the standard ResNet50 backbone \citep{he_deep_2015} to operate on time–frequency cochleagrams using 2D convolutions. The input to the model is a single-channel cochleagram of shape $(1, 211, 390)$, representing 211 frequency bins across 390 time steps. The topographic loss was applied to the second convolutional layer within each residual block, promoting spatial smoothness and topographic organization across successive hierarchical stages of the network (n total = 16 layers). All models were trained on 4$\times$H200 NVIDIA GPUs using identical multi-task objectives (see next section) and hyper-parameters, ensuring a fair comparison across architectures. All hyper-parameters used in pre-training and linear probing of \toponames are in \cref{tab:hyperparameter}


\textbf{Training objective.} All models were trained on the Word-Speaker-Noise dataset \citep{feather2019metamers}, which supports multi-task learning for (1) word recognition, (2) speaker identification, and (3) background noise classification. The dataset includes 230,356 speech clips across 793 word classes and 432 speaker identities, with class sampling designed to reduce overlap (no more than 25\% samples from any one word-speaker pair). Background audio was drawn from 718,625 curated AudioSet clips consisting of human and animal sounds, various musical clips, and environmental sounds to ensure diverse and high-quality noise. Model training sounds included speech-only, noise-only, and speech+noise mixtures, with augmentations such as random cropping, RMS normalization, and variable SNR mixing ($-10\,\mathrm{dB}$ to $+10\,\mathrm{dB}$). 

\subsection{Model evaluations}

We benchmarked the performance of models trained with and without topography across a range of auditory domains: ESC-50 \citep{piczak_esc_2015} for environmental sound classification, NSynth \citep{engel_neural_2017} for musical instrument classification, and Speech Commands \citep{speech_commands} for word and speaker recognition. Lastly, we evaluated these models on human auditory cortex datasets including two fMRI datasets NH2015 \citep{norman2015distinct}, and B2021 \citep{boebinger2021music}, and an ECoG dataset \citep{norman2022neural}.


\subsubsection{Accuracy}

For ESC-50, which includes 2,000 environmental sound clips across 50 categories, we followed the standard five-fold cross-validation protocol. Representations were taken from the penultimate layer of each model (e.g., \texttt{AvgPool} for ResNet50). Each 5-second ESC-50 clip was randomly cropped into five 2-second segments to match the model input duration. All five crops of a training clip were used as independent training samples. During evaluation, we applied majority voting across the five crops of each test clip to determine the predicted class label. We applied cross-validation over five regularization parameters (\( C = [0.01, 0.1, 1.0, 10.0, 100.0] \)). The final accuracy was averaged over the five ESC-50 folds. For NSynth ($\sim$300{,}000 musical notes, 11 instrument families), we used the first 2 seconds of each 4-second clip and extracted final-layer representations. A linear SVM was trained on the official validation split and evaluated on the test split using top-1 accuracy.

For Speech Commands v2, which contains approximately 100,000 1-second utterances of 35 spoken command words (e.g. yes, no, up, down). To ensure consistency with model's input during training, each audio clip was zero-padded to 2-seconds. We extracted frozen representations from the final pooling layer and trained a linear SVM to classify the command labels. We followed the standard validation-test split provided by the dataset: the SVM was trained on the validation set and evaluated on the held-out test set. Accuracy was measured as top-1 accuracy across all 35 classes.

\subsubsection{Brain Predictivity} We evaluated brain predictivity using two human auditory fMRI datasets: NH2015 \citep{norman2015distinct} and B2021 \citep{boebinger2021music}. NH2015 contains responses from 8 participants to 165 two-second natural sounds spanning a diverse set of real-world categories. The stimuli were behaviorally validated using a 10-way Mechanical Turk classification task, retaining sounds with at least 80\% classification accuracy \citep{tuckute_many_2023}. We followed the voxel-selection and preprocessing procedures used in the original model--brain evaluation study.

B2021 \citep{boebinger2021music} contains responses from 20 participants, consisting of 10 highly trained musicians and 10 non-musicians. Its stimulus set includes 192 natural sounds: the 165 sounds from NH2015 \citep{norman2015distinct} and 27 additional music and drumming clips drawn from diverse musical cultures. To ensure a direct comparison between datasets, all analyses in the present study were restricted to the 165 sounds shared by NH2015 and B2021.

To quantify how well model representations predict human auditory responses, we fit voxelwise linear encoding models linking model activations to fMRI responses. For each voxel and model layer, we trained regularized linear (ridge) regressions on time-averaged activations using cross-validation across sound stimuli, and evaluated performance on held-out sounds using Pearson correlation ($r$). To ensure fair comparison across voxels and models, prediction accuracy was corrected for measurement reliability using an attenuation-corrected variance-explained metric. Full details of the regression procedure, cross-validation scheme, and reliability correction are provided in Appendix.

\begin{figure*}[t]
  \centering
  \includegraphics[width=0.9\textwidth]{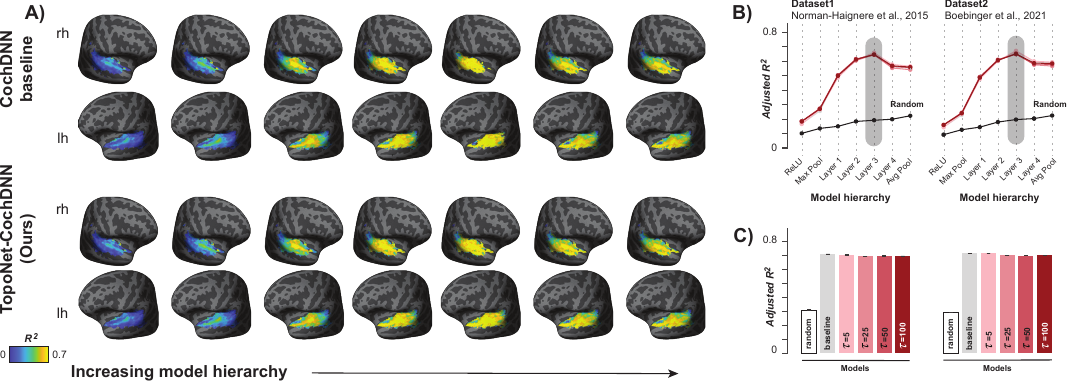}
  \caption{\textbf{Topographic auditory models maintain overall voxel-wise brain predictivity.} 
    A) Brain maps display peak voxel-wise predictions across the auditory cortex for the baseline and topographic ANNs. The colormap indicates the mean variance explained at each cortical vertex, averaged across subjects on the fsaverage surface. The model layers are shown from left to right, corresponding to increasing depth in the network hierarchy. B) ANNs prediction accuracy as a function of depth in the network hierarchy. Note that the baseline model and the \toponames are highly overlapping. The black line is randomly initialized model. The vertical dashed lines represent different layers within the model. C) Bar plot summarizing peak encoding performance across all voxels for each model variant, including the random-initialized, baseline, and \toponames models. 
  }
  \label{fig:fig3}
\end{figure*}

\subsubsection{Features and Voxel Decomposition}

Functional selectivities for speech and music are often not directly observable at the voxel level in fMRI data, and are instead inferred by decomposing voxel responses into a small number of latent components \cite{norman2015distinct}. To analyze the structure of model and brain responses, we adapted this voxel decomposition technique to the fMRI response matrix and model activations. Specifically, we adopted the non-parametric decomposition algorithm \citep{norman2015distinct}, which factorizes a data matrix $D \in \mathbb{R}^{S \times V}$ (where $S$ for stimuli and $V$ for voxels/features) into a response matrix $R \in \mathbb{R}^{S \times N}$ and a weight matrix $W \in \mathbb{R}^{N \times V}$ (where $N$ for number of components), such that:

\begin{equation}
    D \approx R W
\end{equation}

\subsubsection{Greedy Component Matching}


Because the number of inferred components differed between models and human ECoG (fixed at $N = 15$) or fMRI (fixed at $N = 6$), a direct one-to-one comparison was not possible. To align component structures, we adopted a greedy matching procedure based on correlation. For each pair of model and ECoG component sets, we first computed the full correlation matrix between their response profiles across stimuli. For ECoG, this analysis was performed over time, such that component response profiles captured their temporal dynamics across stimulus presentation. Components were then matched iteratively in descending order of correlation magnitude. At each step, the pair with the highest remaining correlation was selected, and both components were removed from further consideration (i.e., matching was performed without replacement). This procedure yields a one-to-one alignment between ECoG/fMRI components and a subset of model components that maximizes overall correspondence while preventing multiple assignments to the same component. The resulting matched pairs were used for all component-level alignment analyses, measured using Pearson correlation ($r$).

\subsubsection{Explained Variance from ICA Decomposition}

To quantify the compactness of model representations, we measured the amount of variance explained as a function of the number of inferred components. After applying the ICA decomposition, we reconstructed the data matrix using the top $k$ components and computed the cumulative explained variance.

We then determined the minimum number of components required to explain $80\%$ of the variance for each model. Comparing this threshold across models provides a measure of intrinsic dimensionality: models requiring fewer components to reach the same explained variance are considered to have more compact and structured internal representations.

\subsubsection{Representational Similarity Analysis (RSA)}

To complement the component-level alignment analyses, we assessed the geometry of model and brain representations using representational similarity analysis \citep{kriegeskorte2008representational}. For each model and each time point in the ECoG recordings, we constructed a representational dissimilarity matrix (RDM) by computing pairwise dissimilarities (1 minus Pearson correlation) across all 165 sounds. Alignment between model and ECoG RDMs was quantified using Spearman rank correlation ($\rho$), which measures the degree to which the pairwise similarity structure of the model matches that of the brain at each moment in time.

To obtain a time-resolved measure of alignment, this procedure was repeated independently for each ECoG time point (sampled at the broadband gamma envelope resolution), yielding a time series of RSA correlations. We report the mean $\rho$ across all ECoG electrodes, with shaded regions indicating variability across electrodes. All RSA analyses were performed on Layer 3 activations, consistent with the encoding model results showing that this layer achieves peak brain predictivity.

\section{Results}

The main goal of this study is to understand how internal component representations differ between baseline and topographic audio ANNs (TopoAudio models). Before addressing this question, we first establish that the TopoAudio models serve as strong candidate models of auditory representations by evaluating both their task performance and their ability to predict neural responses in human auditory cortex measured with fMRI. 

\textbf{\toponames models preserve performance under topographic constraints.}
A typical concern with topographic neural network models is that introducing spatial constraints often leads to substantial drops in task performance on standard engineering tests \citep{qian2024local}. To examine whether this tradeoff holds in the auditory domain, we evaluated \toponames with varying levels of topographic smoothness (controlled by hyperparameter $\tau$), against baseline non-topographic models on three benchmark datasets: environmental sounds, speech, and music (see Methods). All results are obtained from models trained on the same mixed sound dataset. Differences across rows reflect evaluation conditions, not different models. Please see the Appendix for details on how smoothness is computed.

\begin{table}[h!]
\caption{\textbf{Performance under topographic constraints.} Topographic models improve smoothness with minimal changes in accuracy; Topographic Avg. averages non-baseline $\tau$ values.}
\centering
\begin{adjustbox}{width=\linewidth}
\begin{tabular}{lccc c}
\toprule
\textbf{Topography ($\tau$)} & \multicolumn{3}{c}{\textbf{Accuracy}} & \textbf{Smoothness} \\
\cmidrule(lr){2-4}
 & ESC50 & NSynth & SpeechCmd &  \\
\midrule
\multicolumn{5}{l}{\textbf{ResNet-50}} \\
\textcolor{gray}{Baseline} & \textcolor{gray}{81.69} & \textcolor{gray}{98.29} & \textcolor{gray}{86.68} & \textcolor{gray}{-0.002} \\
5        & 81.46 & 98.50 & 86.88 & 0.16 \\
25       & 80.62 & 98.39 & 87.89 & 0.22 \\
50       & 80.84 & 98.36 & 87.57 & 0.50 \\
100      & 80.32 & 98.72 & 86.85 & 0.62 \\
\rowcolor{gray!10} \textit{Topographic Avg.} & 80.79 & 98.49 & 87.33 & 0.38 \\
\bottomrule
\end{tabular}
\end{adjustbox}
\label{table:performance}
\end{table}

\textbf{Table 1} Model performance across multiple evaluation benchmarks (ESC50, NSynth, Speech Commands, etc.) for baseline (gray) and topographic models. While baseline models achieve slightly higher accuracy, introducing topographic constraints ($\tau$) substantially increases representational smoothness with only modest changes in classification performance. Topographic Avg. indicates the mean performance across all non-baseline $\tau$ values.

\paragraph{Topographic auditory models predict human fMRI responses with high accuracy.}
A desirable feature of a biologically inspired model is that it must predict neural responses with high accuracy. To assess whether brain predictivity is maintained even with topographic constraints ($\tau$), we evaluated the predictive abilities of \toponame s and the baseline ResNet50 architecture using two fMRI datasets, NH2015 \citep{norman2015distinct} and B2021 \citep{boebinger2021music}. Following prior work \citep{tuckute_many_2023}, we computed the voxel-wise explained variance (adjusted $R^2$) using linear regression across each layer of the model hierarchy.

\Cref{fig:fig3} shows that both baseline and \toponame s achieved comparable levels of predictivity across layers. This is evident in the qualitative prediction maps projected onto the cortical surface (\Cref{fig:fig3}A), where both models showed similar levels of adjusted $R^2$ across voxels. Both models also showed characteristic features of cortical predictivity. Prediction accuracies improve across model layers. This is apparent more clearly in \Cref{fig:fig3}B which shows that model prediction accuracy peaks around the layer 3. It is particularly striking that prediction accuracy is very similar between the baseline and \toponame s. The five curves corresponding to the baseline model and the different topographic variants of \toponame s are virtually superimposed. As a further quantification, \Cref{fig:fig3}C demonstrates that adding the topographic constraint does little to reduce the overall model prediction accuracy. This pattern holds not only across the entire auditory cortex but also within specific regions of interest. When we repeated the analysis separately for early tonotopic regions, pitch-selective regions, and higher-order music and speech areas, we observed similar trends. \toponame s were as predictive as baseline models for all ROIs. Together these results confirm that inducing topographic constraints to audio models does not impair their ability to predict fMRI responses in human auditory cortex. Thus, \toponame s preserve both task performance (on engineering metrics) and predictive accuracy (on 2 distinct fMRI datasets) while producing spatially structured internal representations. The similarity of the TopoAudio to the baseline models leads to a natural question: \textbf{are these models actually the same, or are our evaluations not sensitive enough to reveal the differences?}

\begin{figure*}[t]
  \centering
  \includegraphics[width=\textwidth]{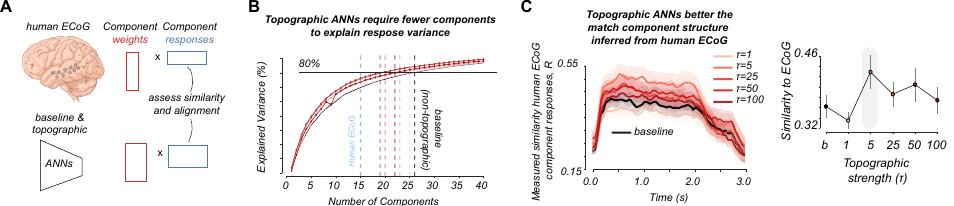}
  \caption{\textbf{A.} Schematic of the component analysis and alignment procedure. Human ECoG responses and model activations are decomposed into latent components using a non-parametric ICA approach. Component response profiles are then aligned across brain and model using correlation-based matching to assess structural similarity. \textbf{B.} Topographic ANNs require fewer components to explain response variance. Cumulative explained variance as a function of the number of components shows that models with topographic constraints reach 80\% variance with fewer components than the non-topographic baseline, approaching the dimensionality observed in human ECoG. The dashed vertical lines represent the number of components required to reach 80\% explained variance in topographic (red) and baseline (black) models. \textbf{C.} Topographic ANNs better match the component structure inferred from human ECoG. Correlation time courses between matched model and ECoG components reveal stronger alignment for topographic models compared to the baseline. The summary (right) shows mean similarity across components, indicating improved brain–model correspondence at intermediate levels of topographic strength.}
  \label{fig:summary}
\end{figure*}

\paragraph{Topographic constraints produce more compact component structure.}
We next examined how information is organized within model representations and whether imposing topographic constraints alters this structure relative to standard (baseline) audio networks. In human auditory cortex, neural responses have been shown to decompose into a set of components \citep{norman2015distinct}. Many of these components are interpretable and have been linked to neural response patterns selective for speech, music \citep{norman2015distinct, boebinger2021music}, and, more recently, song \citep{norman2022neural}. Note that the number of inferred dimensions depends on the measurement resolution: fMRI studies typically recover a relatively small number of components (e.g., $N \approx 6$, \citep{norman2015distinct}), whereas higher-resolution ECoG recordings recover more (e.g., $N \approx 15$, \citep{norman2022neural}). Both numbers reflect the specific stimulus sets and decomposition procedures used in those studies; we adopt the same stimuli and algorithm to ensure direct comparability. Here, we ask whether topographic models better match those observed component structures compared to baseline models.

To this end, we performed an ICA component analysis on model activations (analogous to the brain). To remove additional experimental degrees of freedom, we focused only on activations from Layer 3 (the layer that best predicts neural responses). We found that increasing the strength of topography (i.e. increasing $\tau$) led to a systematic reduction in the number of inferred components (e.g., $N=26$ for the baseline model vs. $N=20$ for $\tau=25$) indicating that topographic organization yields more structured and compact representations. Note that this general pattern is expected from prior work that has shown that imposing topographic constraints reduces the intrinsic dimensionality of model representations \citep{margalit_unifying_2024, qian2024local, deb2025toponets}. What remains unclear, and could not be tested in prior work, is whether the components inferred from topographic models \textit{better} align with the components inferred from human ECoG recordings. We tested this question next. 

\textbf{Topographic constraints yield more brain-aligned component structure.}
Following \textbf{Fig. 3B} (and prior work \citet{norman2015distinct}), we selected the optimal number of components required to reach 80\% explained variance. We then aligned model and ECoG components using a greedy matching procedure (see Methods) and compared the inferred components from ANN models with the timecourse of the 15 components derived from ECoG recordings in the brain. The mean correlation between inferred model components and ECoG component responses, averaged across all 15 ECoG components, is shown in \textbf{Fig. 3C}. We find that models with topographic constraints (shades of red lines) consistently showed higher correlations with ECoG components than the baseline model (black line). Topographic models maintained high correlations throughout the stimulus window (0–3 s). The difference between the baseline and topographic models was statistically significant across 15 components (at $\tau=5,50$, P<0.05, P=0.07 for $\tau=25$, Fisher z posthoc paired t-tests). A possible reason for this pattern is that topographic models just have fewer components than baseline models. To rule this out, we repeated the analysis using a fixed number of components (25, the maximum inferred for the baseline model) across all models. All our results remain qualitatively unchanged (See Appendix). Together, these results provide initial evidence that topographic constraints, on average, yield representational components that better aligned with the component structure derived from human ECoG recordings. To further corroborate this finding using a geometry-based measure, we computed representational similarity analysis (RSA) between model and ECoG representations over time. As shown in \Cref{fig:rsa_time}, topographic models consistently achieve higher RSA correlations with ECoG throughout the stimulus window, with alignment peaking at intermediate topographic strengths ($\tau=5$--$25$) and remaining above baseline for larger $\tau$.

\begin{figure*}[t]
  \centering
  \includegraphics[width=\textwidth]{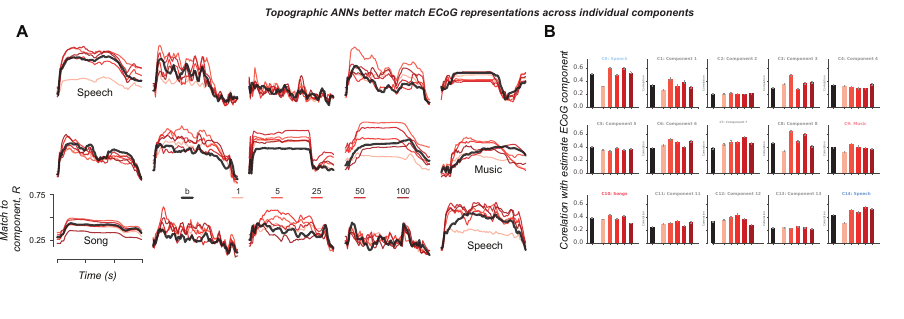}
  \caption{\textbf{Topographic ANNs better match ECoG representations across individual components.} 
\textbf{A.} Correlation time courses between each model-derived component and its matched human ECoG component. Each panel shows the matched component pair, with interpretable ECoG components (e.g., speech, music, song) labeled. Red lines denote topographic models at increasing topographic strengths ($\tau$), and black denotes the non-topographic baseline. 
\textbf{B.} Mean similarity collapsed across time for each matched component. Bars show the average correlation between model and ECoG component responses.}
  \label{fig:summary}
\end{figure*}

Next, we examined the alignment between the ECoG component and our baseline and \toponame s on a per-component basis. The correlation time course between each model-derived component and its matched ECoG component is shown in Figure 4A. Prior work has shown that several of these ECoG components are interpretable and correspond to response patterns selective for speech, music, and song. The interpretable components are marked in Figure 4A. Across individual components, topographic models were typically at par with, and more often outperformed, the baseline model. We summarize these effects by collapsing correlations across time in Figure 4B. In general, a modest topographic strength ($\tau=5$) performed best, often outperforming models with higher topographies, suggesting a sweet spot in which topographic constraints improve brain alignment without over-constraining the representation.

Finally, we zoomed in on the components on the speech, music, and song components that are most directly interpretable. Figure 5 shows the category response profile for the ECoG components (center panel), the baseline model (left panel) and a topographic audio model ($\tau=5$, right panel). Across these components, the topographic model recovers the ECoG-derived tuning profiles more faithfully than the baseline (other component responses are presented in \Cref{fig:ica_weights_visualization_baseline,fig:ica_weights_visualization_tau5}).  For example, the speech components (C1 and C15) show strong positive responses for speech categories in ECoG. The topographic model reproduces this selectivity more cleanly, whereas the baseline shows weaker and more mixed responses across categories. Similarly for the music component (C10), the ECoG component has a distinct preference for music and the topographic model captures this preference with a closer match in both sign and relative magnitude relative to the baseline model. In contrast, the song component (C11) is less cleanly recovered. Although ECoG shows an interpretable song-related profile, both models show weaker and less specific selectivity, suggesting that this component is harder to capture with current model representations. As expected, the improvement in component alignment is accompanied by more spatially contiguous maps in the topographic model (component weight maps, Fig. 5, right). Units contributing to each component appear spatial clustered in topographic model compared to baseline model. Together these examples show topographic models yield model components with clear, interpretable tuning profiles that more closely mirror the component structure inferred from human ECoG. 

\section{Discussion}
In this work, we asked whether imposing topography in audio ANNs (\toponame) changes the \textit{internal} representations of the model to better match the brain's component structure. We first established topographic models as viable candidate models of auditory processing, showing they achieve task performance comparable to baseline models and predict human auditory fMRI responses at a similar level. Despite matched performance on these standard evaluations, topographic constraints systematically reduced the number of inferred representational components. More importantly, the components inferred from topographic models align \textit{more} closely with those derived from human ECoG recordings, both on average (univariate and RSA) and at the level of individual interpretable components linked to speech, music, and song. Together, these results provide evidence that topographic constraints reshape internal population codes in a brain-aligned manner even when standard performance metrics are matched.
\begin{table}[t]
\centering
\small
\setlength{\tabcolsep}{6pt}
\begin{tabular}{lcc}
\toprule
\rowcolor{gray!12}
Metric & Baseline & Topographic Audio \\
\midrule
Engineering tests & $\checkmark$ & $\checkmark$ \\
Brain prediction & $\checkmark$ & $\checkmark$ \\
Smoothness & $\times$ & $\checkmark$ \\
ICA components & more & fewer \\
Component alignment & worse & better \\
\bottomrule
\end{tabular}

\caption{\textbf{Baseline v. topographic audio models.} Both models achieve comparable task performance and fMRI predictivity, but topographic models learn smoother representations, fewer ICA components, and stronger alignment with brain-derived components, indicating more brain-like internal structure.}
\label{tab:model_comparison}
\end{table}

A key contribution of this work is the \toponames models themselves. While topographic constraints have been explored in vision, and more recently in language, topographic auditory networks have been largely missing from the toolkit of auditory neuroscience. Here we provide a first suite of trained \toponames models. These models retain strong task performance and neural predictivity and can offer a concrete, reusable starting point for studying how topography emerges in audio models. At the same time we focus here on one strategy for inducing topography into ANN models. An important direction for future work is to compare alternative mechanisms for enforcing spatial organization and ask whether particular forms of topographic constraint yield more brain-aligned auditory representations. A major contribution of this work is the focus on component structure of the internal representation as a consequence of topography. Much of the prior literature on topographic models has emphasized the \textit{spatial appearance} of maps. And while it has been reported that topographic constraints reduce representational dimensionality, it has been hard to ask the more mechanistic question: do the emergent components actually correspond to the brain’s own representational ``parts list''?  The auditory domain made this test possible because the component structure of auditory cortex has been characterized across both fMRI and ECoG.

The fact that we observe component-level alignment is consequential. It suggests that topographic constraints actively bias internal representations toward a regime the brain itself uses. This sits naturally alongside long-standing ideas from efficient coding \citep{barlow1961possible} and sparse coding \citep{olshausen1996emergence}, which show that sensory systems organize representations to be compact, minimally redundant, and decomposable into meaningful factors \citep{mesgarani2014mechanisms}. Our results point to a plausible mechanistic link between spatial constraints and decomposable population codes: when nearby units are encouraged to be similar, the network is pushed toward locally redundant, smoothly varying representations. These results call for more systematic theoretical work to formalize when topography should be expected to yield compact codes, what trade-offs it imposes on task demands, and which forms of topography best predict brain components across recording modalities.

\begin{figure}[t]
  \centering
  \includegraphics[width=\linewidth]{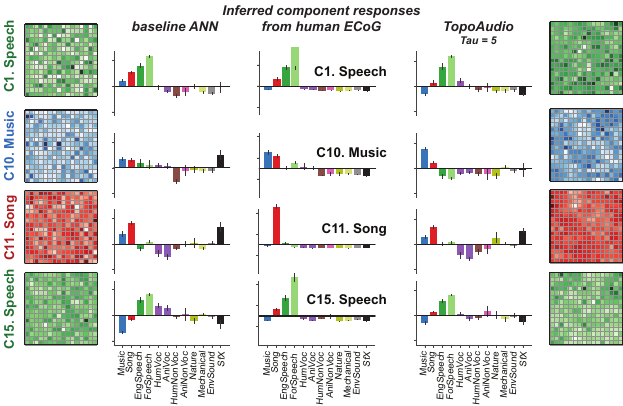}
  \caption{\textbf{Topographic models better recover interpretable ECoG component tuning and spatial structure.}
Category response profiles for speech (C1, C15), music (C10), and song (C11) components are shown for human ECoG (center), the baseline ANN (left), and a topographic model ($\tau=5$, right). The topographic model more faithfully reproduces the speech and music selectivity observed in ECoG, capturing cleaner category preferences than the baseline. The song component is less well recovered by both models. Corresponding weight maps show more spatially contiguous unit organization in the topographic model.}
  \label{fig:summary}
\end{figure}

Our analyses also identified model components without clear counterparts in human ECoG (yet). One possibility is that these components may also exist in the brain but remain undetected due to electrode sampling biases or resolution limits. Follow-up experimental work can now directly look for them using targeted stimuli and fMRI contrasts. The song component was also less cleanly recovered in our models, suggesting higher-level auditory structure may need richer training signals or different model architectures. Future work should test robustness to alternative decomposition methods (like NMF, \cite{khosla2022highly}) and examine how component structure evolves across network depth. Together with recent work on topographic constraints in vision and language, these results point to a domain-general principle: wiring-length constraints may be a general mechanism for producing compact, interpretable internal representations across sensory systems. The fact that the same inductive bias reshapes population codes in vision, language, and now audition is striking. It suggests that topography is a fundamental organizing principle of the brain. We hope TopoAudio provides a concrete foundation for testing this idea further, and that component-level alignment across domains becomes a richer benchmark for evaluating topographic models going forward.

Overall, these results suggest that topographic constraints in audio models reshape internal representations toward brain-derived component organization. Thus topography might be a useful inductive bias for building brain-aligned ANNs and that component-level alignment could serve as an alternate benchmark for topographic models.

\begin{figure}[t]
  \centering
  \includegraphics[width=0.9\linewidth]{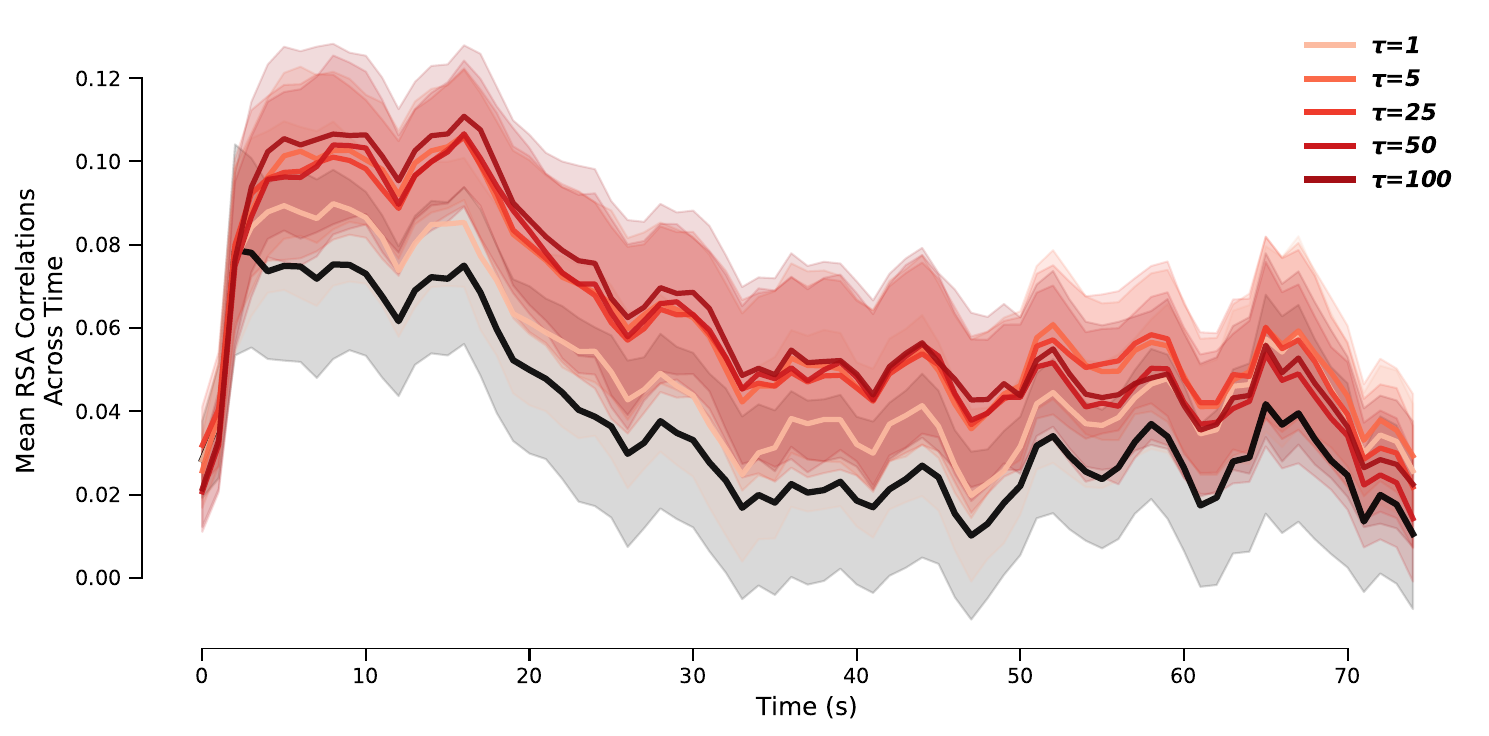}
  \caption{\textbf{RSA between model representations and ECoG responses over time (Layer 3).} Mean Spearman correlation ($\rho$) between model and ECoG representational dissimilarity matrices, plotted across the stimulus window. Shaded regions indicate variability across components. Topographic models (shades of red) consistently outperform the non-topographic baseline (black), with peak alignment at intermediate topographic strengths ($\tau=5$--$25$).}
  \label{fig:rsa_time}
\end{figure}

\section{Limitations}

While our results provide initial evidence that topographic constraints yield more brain-aligned internal representations in auditory models, several limitations should be noted.

\textbf{Single architecture.} All \toponames models are built on the CochResNet50 backbone. It remains to be seen whether the observed benefits of topographic constraints generalize to other architectures, such as transformer-based audio models (e.g., AST). Testing across architectures is important for establishing the robustness of the reported effects.

\textbf{Single topographic strategy.} We employed one particular form of topographic regularization (TopoLoss via weight blurring, following \citet{deb2025toponets}). Alternative mechanisms — such as distance-dependent response similarity constraints \citep{lee2020topographic, margalit_unifying_2024} or lateral interaction-based approaches \citep{qian2024local} — may yield different component structures or stronger brain alignment and warrant systematic comparison.

\textbf{Stimulus set.} All analyses are based on 165 natural sounds from a single validated stimulus set \citep{norman2015distinct}. The inferred components, and thus the alignment results, may be sensitive to the choice of stimuli. Future work should test whether the observed improvements in component alignment hold for larger and more diverse sound sets.

\printbibliography
\clearpage

\section{Appendix}

\subsection{Voxelwise Response Modeling}
\label{sec:voxelwise_modeling}
The voxel-wise procedure was repeated 10 times (once per train-test split), and the median variance explained (noise-corrected) was reported for each voxel-layer pair. We evaluated all layers from each candidate model on both datasets, yielding voxelwise explained variance values for 7,694 voxels (NH2015) and 26,792 voxels (B2021). 

\paragraph{Regularized linear regression and cross-validation.}
To model the relationship between model unit activations and measured brain responses, we used voxelwise linear encoding models. For each voxel, we predicted its time-averaged response to natural sounds as a linear combination of time-averaged activations from a specific model layer. We randomly split the 165 sounds into 10 unique train-test partitions of 83 training and 82 test sounds. For each split, we fit a regularized linear regression (ridge regression) model using the 83 training sounds and evaluated prediction performance on the held-out 82 sounds.

\paragraph{Regression formulation.}
Let $\mathbf{y} \in \mathbb{R}^n$ be the voxel’s mean response to $n = 83$ sounds, and let $\mathbf{X} \in \mathbb{R}^{n \times d}$ be the matrix of $d$ regressors (i.e., time-averaged activations from a model layer). The ridge solution is:

\begin{equation}
\mathbf{w}
=
\left(
\mathbf{X}^\top \mathbf{X}
+
n \lambda \mathbf{I}
\right)^{-1}
\mathbf{X}^\top \mathbf{y}
\label{eq:ridge_solution}
\end{equation}

where $\lambda$ is the regularization parameter and $\mathbf{w}$ is the vector of regression weights. Both $\mathbf{y}$ and the columns of $\mathbf{X}$ were demeaned (but not normalized) prior to regression. This allowed units with greater magnitude variance to contribute more to the prediction under a non-isotropic Gaussian prior. To avoid data leakage, all transformations were learned on the training set and applied to the test set.

We used leave-one-out cross-validation within the 83 training sounds to select $\lambda$. For each of 100 logarithmically spaced values (from $10^{-50}$ to $10^{49}$), we computed the mean squared error of the predicted response for each left-out training sound. The $\lambda$ minimizing this error was used to retrain the model on all 83 training sounds. The final model was then used to predict responses to the 82 held-out test sounds, and performance was quantified using the Pearson correlation between predicted and actual voxel responses. Negative correlations or correlations with zero variance were set to zero.

\paragraph{Correcting for reliability of predicted voxel responses.}
Because both training and test responses are affected by measurement noise, we corrected for the reliability of both the predicted and measured voxel responses. This correction was essential to fairly compare model performance across voxels and model layers. We defined the corrected variance explained using the attenuation-corrected squared correlation:

\begin{equation}
    r^2_{\mathbf{v},\hat{\mathbf{v}}}{}^* = \frac{r(\mathbf{v}_{123}, \hat{\mathbf{v}}_{123})^2}{r'_v r'_{\hat{v}}}
\end{equation}

where $\mathbf{v}_{123}$ is the voxel response to the 82 test sounds, $\hat{\mathbf{v}}_{123}$ is the predicted response, and $r'_v$, $r'_{\hat{v}}$ are the reliabilities of the measured and predicted responses, respectively. Reliability was estimated via median Spearman–Brown corrected correlations across scan pairs. For stability, we excluded voxels for which $r'_v$ or $r'_{\hat{v}}$ was less than $k = 0.182$ and $k = 0.183$, respectively (corresponding to $p < 0.05$ thresholds for 83- and 82-dimensional Gaussian variables).

\subsection{Moran's I spatial autocorrelation}

Moran’s I is a metric of global spatial autocorrelation that quantifies the degree to which values defined over a spatial domain exhibit smooth clustering or random dispersion \citep{rathi2024topolm}. Formally, for a set of values $x_i$ arranged over $N$ spatial units, Moran’s I is defined as:

\begin{equation}
I = \frac{N}{W} \,
\frac{\sum_{i}\sum_{j} w_{ij}(x_i - \bar{x})(x_j - \bar{x})}
{\sum_{i}(x_i - \bar{x})^2},
\end{equation}

where $\bar{x}$ is the global mean, $w_{ij}$ are entries of a sparse adjacency matrix encoding the neighborhood relationships among vertices or units, and $W = \sum_{i,j} w_{ij}$ is the total connection weight. Values of Moran’s I range from $-1$ to $1$, with positive values reflecting spatially smooth and contiguous organization, values near zero indicating spatial randomness, and negative values indicating systematic spatial dispersion.



\subsection{Moran's I for Computational Models}

For neural network models, we applied an analogous spatial autocorrelation procedure by imposing a fixed two-dimensional topographic layout over the units of each layer. Specifically, units within each MLP block (or convolutional feature map) were assigned to positions on a 2D grid, and a binary adjacency matrix was constructed based on local grid connectivity. Using this spatial structure, we computed selectivity maps for each model, including frequency tonotopy maps, modulation-rate maps, and music- vs.-other and speech- vs.-other contrast maps derived from model activations. Moran’s I was computed for each layer individually and then averaged across relevant layers to obtain a model-level smoothness estimate for each topographic strength $\tau$. This procedure mirrors the approach used for fMRI maps, enabling direct comparison between cortical organization and the representational smoothness learned by neural networks trained with and without topographic constraints.

\begin{figure*}[t]
  \centering
  \includegraphics[width=\textwidth]{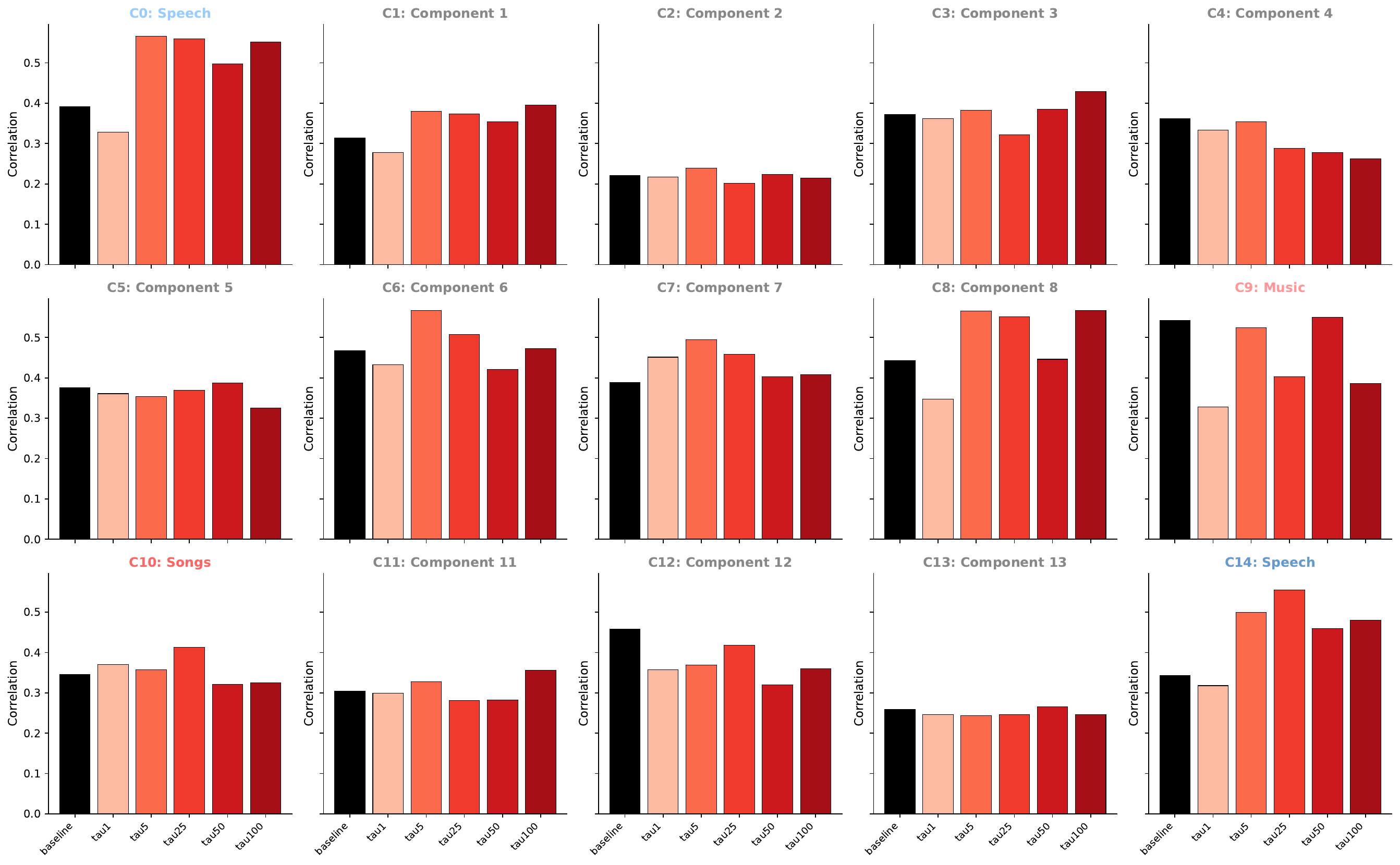}
  \caption{\textbf{Component-level alignment with fixed ICA dimensionality ($N=25$).}
ICA was performed with a fixed number of components ($N=25$) for all models. Bars show the mean correlation (collapsed across time) between each matched model component and its corresponding human ECoG component. Results are shown for the baseline and topographic models across topographic strengths. Even under matched dimensionality, topographic models maintain stronger or comparable component-level alignment relative to the baseline, indicating that improvements are not solely driven by differences in inferred component number.}
  \label{fig:summary}
\end{figure*}

\subsection{ECoG Data}

ECoG data used in this study were obtained from \citet{norman2022neural}. The dataset comprises intracranial recordings from 15 epilepsy patients (mean age $\approx$ 35 years) undergoing clinical monitoring with subdural electrode arrays. Electrode placement was determined by clinical needs, resulting in sparse and heterogeneous cortical coverage, with a focus on auditory regions, particularly the superior temporal gyrus.

Neural activity was recorded using high-density electrode grids (typically 2.3 mm diameter, 6 mm spacing), with some subjects implanted with higher-resolution grids (1 mm diameter, 3 mm spacing). The analysis focused on broadband gamma responses (70–140 Hz), which are known to correlate with local neuronal population activity. Signals were preprocessed using common-average referencing, notch filtering to remove line noise, and bandpass filtering to extract gamma power, followed by envelope computation and normalization relative to pre-stimulus baseline.

Participants were presented with a set of 165 natural sounds (each 2 seconds long), encompassing diverse categories including speech, music (instrumental and vocal), environmental sounds, and other everyday auditory stimuli. Sounds were RMS-normalized and delivered in randomized sequences across multiple runs, with brief inter-stimulus intervals. A subset of sounds was repeated to assess response reliability, and electrodes were selected based on split-half reliability of their gamma responses (correlation > 0.2), yielding 272 sound-responsive electrodes across subjects.

For each stimulus, neural responses were analyzed within a time window aligned to stimulus onset (typically 0 -- 3 seconds). The resulting dataset consists of time-resolved population activity across electrodes, capturing both high temporal resolution dynamics and category-selective auditory responses. The dataset provides precise measurements of neural activity in human auditory cortex, enabling fine-grained analysis of auditory representations beyond what is possible with non-invasive methods.

\begin{figure*}[t]
  \centering
  \includegraphics[width=\textwidth]{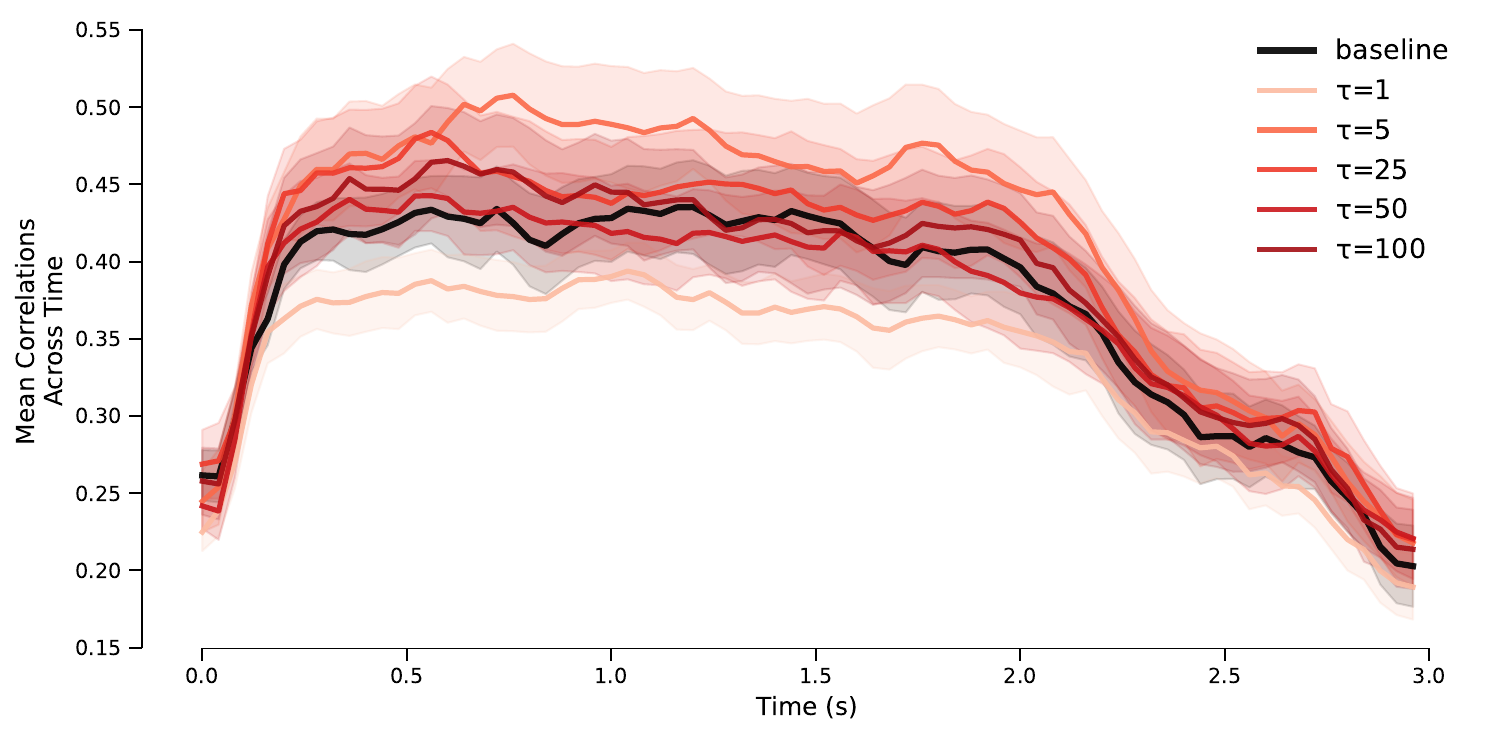}
  \caption{\textbf{Component alignment with fixed ICA dimensionality ($N=25$).}
Correlation time courses between matched model and ECoG components reveal stronger alignment for topographic models compared to the baseline. ICA was performed with a fixed number of components ($N=25$) for all models. Even under matched dimensionality, topographic models—particularly at moderate topographic strength—maintain improved or comparable alignment relative to the baseline, indicating that the observed brain–model correspondence is not driven by differences in inferred component number.}
  \label{fig:summary}
\end{figure*}

\begin{figure*}[t]
  \centering
  \includegraphics[width=\textwidth]{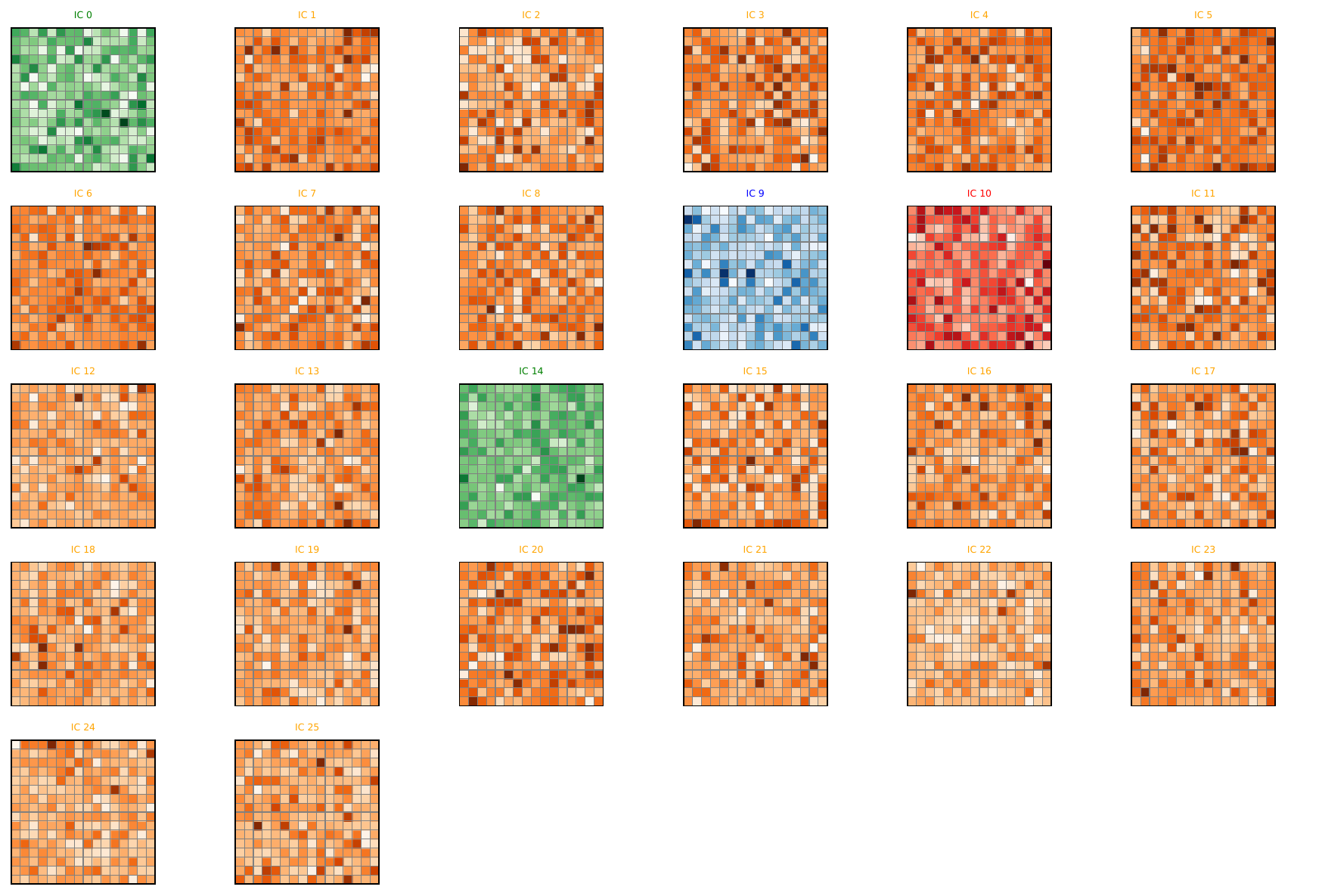}
  \caption{\textbf{Spatial distribution of ICA component weights in the baseline model.}Visualization of all inferred ICA component weight maps projected onto the 2D model grid for the non-topographic baseline. Each panel shows the spatial weight distribution of a single component. Unlike topographic models, component weights in the baseline are spatially diffuse and lack clear clustering, reflecting the absence of imposed spatial organization.}
  \label{fig:ica_weights_visualization_baseline}
\end{figure*}

\begin{figure*}[t]
  \centering
  \includegraphics[width=\textwidth]{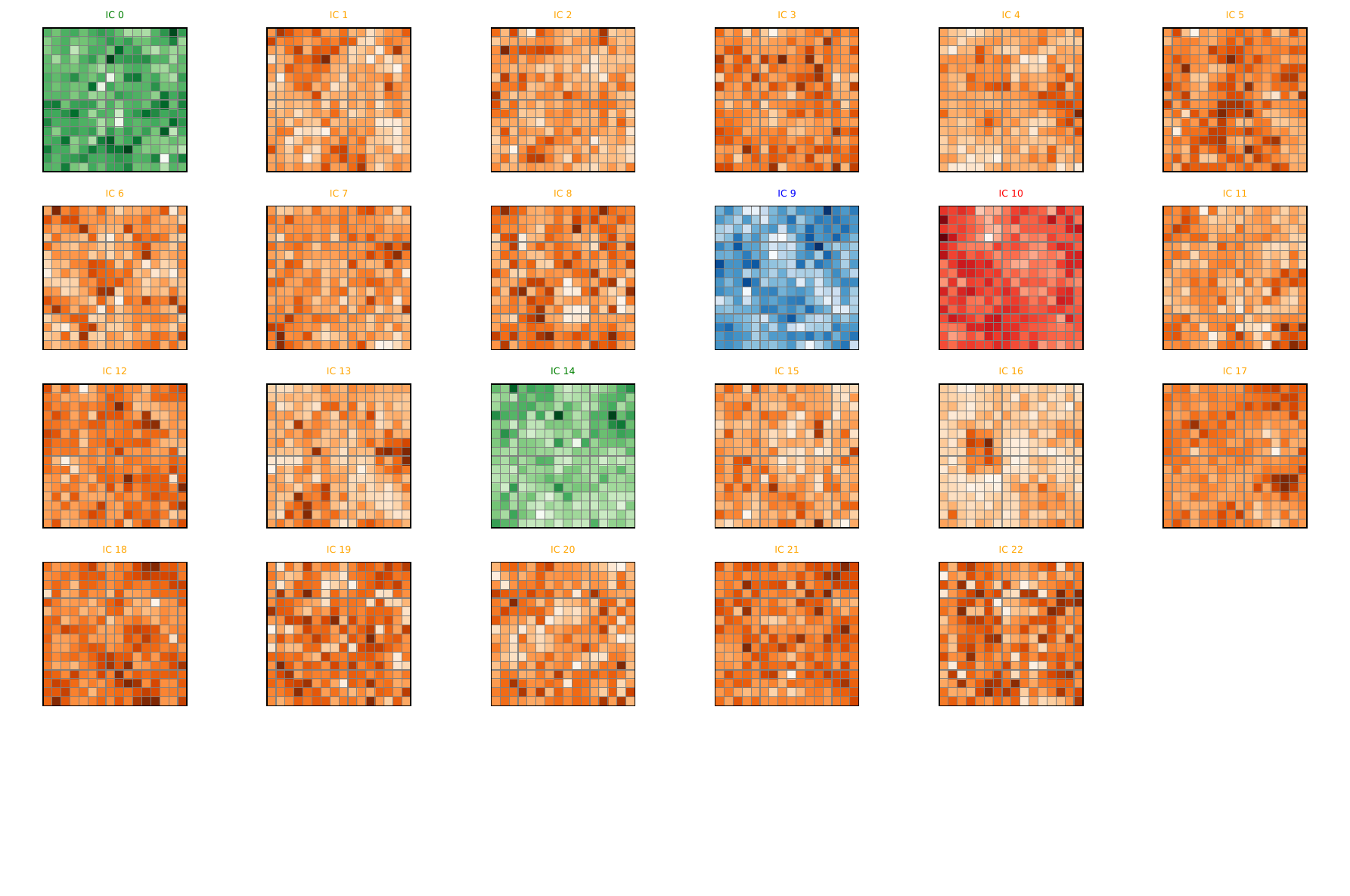}
  \caption{\textbf{Spatial distribution of ICA component weights in the topographic model ($\tau = 5$).} Visualization of all inferred ICA component weight maps ($N=25$) projected onto the 2D model grid for the topographic model with moderate topographic strength ($\tau = 5$). Each panel shows the spatial weight distribution of a single component. In contrast to the baseline model, component weights exhibit clearer spatial clustering and contiguity, reflecting the emergence of organized topographic structure under spatial smoothness constraints.}
  \label{fig:ica_weights_visualization_tau5}
\end{figure*}

\begin{table}[t]
\centering
\small
\caption{Training hyperparameters for TopoAudio.}
\label{tab:hyperparameter}
\begin{tabular*}{\columnwidth}{@{\extracolsep{\fill}} lc}
\toprule
\textbf{Hyperparameter} & \textbf{Training} \\
\midrule
Optimizer        & AdamW \\
Learning rate    & 1e-3 \\
LR scheduler     & Cosine \\
Warmup epochs    & 1 \\
Max epochs       & 6 \\
Batch size       & 256 \\
Weight decay     & -- \\
Betas            & (0.9, 0.999) \\
Momentum         & 0.9 \\
Nesterov         & -- \\
Gradient clip    & 1.0 \\
Precision        & bf16 \\
Random seed      & 1337 \\
\bottomrule
\end{tabular*}
\end{table}

\begin{figure*}[t]
  \centering
  \includegraphics[width=\textwidth]{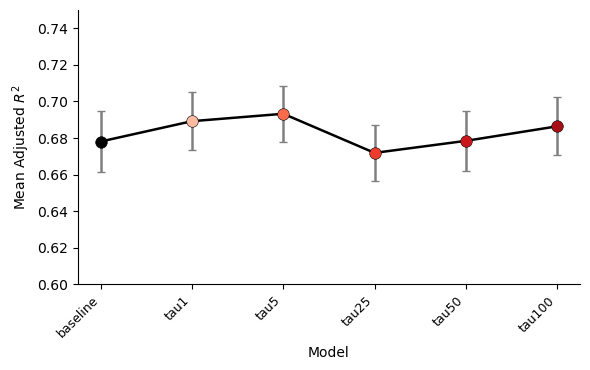}
  \caption{Mean adjusted $R^2$ for predicting raw ECoG electrode responses across models. Performance is averaged over time, with error bars indicating variability across electrodes. Models with intermediate $\tau$ values (e.g., $\tau=1$–$5$) achieve the highest predictive accuracy, while both smaller and larger $\tau$ values show slightly reduced performance. To predict raw ECoG electodes we follow similar procedure as Voxelwise Response Modeling.}
  \label{fig:summary}
\end{figure*}


\end{document}